\documentclass[aps,prl,preprint,groupedaddress]{revtex4}
\usepackage{amssymb,amsmath,amsbsy}
\usepackage{mathrsfs}
\usepackage[dvips]{graphics}
\usepackage{epsfig}

\def\beq{\begin{eqnarray}}
\def\eeq{\end{eqnarray}}
\def\bea{\begin{eqnarray*}}
\def\eea{\end{eqnarray*}}
\def\fig#1{Fig.~\ref{#1}}

\def\tilde{\widetilde}

\def\centeron#1#2{{\setbox0=\hbox{#1}\setbox1=\hbox{#2}\ifdim
\wd1>\wd0\kern.5\wd1\kern-.5\wd0\fi
\copy0\kern-.5\wd0\kern-.5\wd1\copy1\ifdim\wd0>\wd1
\kern.5\wd0\kern-.5\wd1\fi}}
\def\ltap{\;\centeron{\raise.35ex\hbox{$<$}}{\lower.65ex\hbox{$\sim$}}\;}
\def\gtap{\;\centeron{\raise.35ex\hbox{$>$}}{\lower.65ex\hbox{$\sim$}}\;}

\def\lsim{\mathrel{\ltap}}

\newcommand{\newc}{\newcommand}
\newc{\qbar}{{\overline q}}
\newc{\Kahler}{K\"ahler }
\newc{\deltaGS}{\delta_{\rm GS}}

\begin{document}
\preprint{
\vbox{\vspace*{2cm}
      \hbox{SCIPP-08/04}
      \hbox{arXiV:0805.2944~[hep-ph]}
      \hbox{May, 2008}
}}
\vspace*{3cm}

\title{Implementing General Gauge Mediation}
\author{Linda M. Carpenter}
\author{Michael Dine}
\author{Guido Festuccia}
\author{John D.~Mason}
\affiliation{Santa Cruz Institute for Particle Physics and department of Physics  \\
   University of California, Santa Cruz, CA 95064, U.S.A. \\
\vspace{1cm}}

\begin{abstract}
Recently there has been much progress in building models
of gauge mediation, often with predictions different than
those of minimal gauge mediation.  Meade, Seiberg, and
Shih have characterized the most general
spectrum which can arise in gauge mediated
models.    We discuss some of the
challenges of building models of General
Gauge Mediation, especially the problem
of messenger parity and issues connected with
R symmetry breaking and CP violation.  We build
a variety of viable, weakly coupled models
which exhibit some or all of the possible low energy
parameters.
\end{abstract}


\maketitle

\section {Introduction}
\label{introduction}

If evidence for low energy supersymmetry is discovered at the Tevatron or LHC, the most
urgent task will be elucidating the superpartner spectrum.
Gauge mediation is a leading candidate
for the messenger of supersymmetry breaking.  Gauge mediation has a number of virtues:
it is flavor blind, and so accounts for the absence of flavor-changing processes at
very low energies;
it fits well with ideas about dynamical supersymmetry breaking; and, at least in its
simplest forms, it is a highly predictive framework.

Gauge mediation is a well-studied subject, but two developments over the last
few years have lead to a renewal of interest.  First has been the recognition
that theories in which supersymmetry is broken in {\it metastable} ground states,
are common, even generic\cite{iss} \footnote{In many earlier models, both
weakly and strongly interacting, the inclusion of messengers lead to the existence
of vacua in which supersymmetry was restored, but the possibility that the hidden
sector, by itself, was described by a metastable, supersymmetry breaking state,
had not received great attention.}.
This has greatly enlarged -- and simplified -- the possibilities for model building\cite{dfs,dm}.
Second has been the growing appreciation that supersymmetric models generally --
and gauge mediated models in particular -- must be tuned if they are to yield
electroweak symmetry breaking with Higgs particles and superpartners consistent
with experimental constraints.

Early models of gauge mediation did not invoke dynamical supersymmetry breaking.
In their simplest version, so-called {\it minimal gauge mediation} (MGM), there was
a singlet field, $X$, coupled to a set of messengers, filling out a $5$ and
$\bar 5$ representation of $SU(5)$\cite{dinefischler,earlymgm}.  The vacuum
expectation value of the field
$X$ had the form
\beq
\langle X \rangle = x + \theta^2 F_x
\eeq
with couplings to messengers:
\beq
\label{simplemgm}
W = X\left(\lambda_q q \tilde q + \lambda_\ell \ell \tilde \ell\right).
\eeq
The resulting sfermion spectrum is easily calculated:
\beq
\widetilde m^2 ={2 \Lambda^2}
\left[
C_3\left({\alpha_3 \over 4 \pi}\right)^2
+C_2\left({\alpha_2\over 4 \pi}\right)^2
+{5 \over 3}{\left(Y\over2\right)^2}
\left({\alpha_1\over 4 \pi}\right)^2\right],
\label{scalarsmgm}
\eeq
where $\Lambda = F_x/x$.
$C_3 = 4/3$ for color triplets and zero for singlets,
$C_2= 3/4$ for
weak doublets and zero for singlets.
This formula predicts definite ratios of squark,
slepton and gaugino masses.
In the early models, O'Raifeartaigh-like couplings of $X$ to various fields were responsible
for $F_x$.  $R$ symmetries were typically broken explicitly, and so the models were not
natural in the strictest sense.

The first well-studied models of dynamical supersymmetry breaking (DSB)
possessed stable, non-zero energy vacua, before coupling
to messengers.  It was possible
to construct models of minimal gauge
mediation with such DSB, but the resulting theories were quite
baroque\cite{dinenelson}.
The usual strategy was to invoke one (or more!) hidden sectors, whose couplings
to singlets like $X$ eventually yielded the desired structure.  Typically,
once messengers were included, the desired vacuum
state was metastable.  But only with the observation of
of Intriligator, Shih and Seiberg that metastable, dynamical
supersymmetry breaking is a generic
phenomenon, was it appreciated that
one should consider hidden sectors in which
(even before coupling to messengers or MSSM fields) the supersymmetry-breaking
vacuum is metastable.  This
has greatly expanded the possibilities for model building\cite{iss}.

Even in this broader framework, there are model-building challenges.  For models
with stable vacua, there is a theorem due to Nelson and Seiberg that the underlying
theory, if {\it generic}, must possess an R symmetry\cite{nelsonseiberg}.
In the metastable models,
this requirement is relaxed to the requirement of an approximate R symmetry.
This would seem an advantage for model building, since one does not have to account
for spontaneous breaking of the symmetry.  In the simplest ISS model, however, the low
energy theory has an accidental, unbroken R symmetry.  In the ``retrofitted" models
of \cite{dfs}, the simplest models also have such a symmetry at the level of the low
energy effective theory.  Shih\cite{shih} has formulated a general theorem which suggests
that this problem is challenging:  in a weakly coupled theory without gauge interactions
and with a global continuous R symmetry,
the $R$ symmetry is never spontaneously broken if the fields all have R charge $0$
or $2$ (as in the simplest O'Raifeartaigh models).  This result does not hold
in the presence of gauge interactions\cite{dm,iss2}, but the simplest models which
exploit this loophole are not
particularly attractive and are sometimes fine-tuned. This theorem can be circumvented, as shown in \cite{katz}, when a hierarchy between masses can lead to two-loop effects that dominate the one-loop Coleman-Weinberg potential.  Shih has exhibited models with
fields with more exotic R charges (and without
gauge interactions) in which the R symmetry is spontaneously broken.

But, as we will comment further below, these models often introduce new
challenges.  The authors of \cite{dm2} constructed retrofitted models without low
energy R symmetries altogether.  But these models required small couplings which,
while technically natural, added additional complexity, and also raised the value
of the underlying supersymmetry breaking scale.

Another challenge for gauge-mediated models involves the MGM spectrum itself.
Eqn. [\ref{scalarsmgm}] predicts definite ratios of squark,
slepton and gaugino masses.  Coupled with the current limits on the lightest
sleptons (approximately $100$ GeV), it implies that
slepton doublets have masses greater than $215$ GeV, while squark masses are larger than
$715$ GeV.  If one considers
the top/stop contribution to $m_{H_U}^2$ (the coefficient of
$\vert H_U \vert^2$ in the lagrangian),
this is logarithmically divergent.
The cutoff, $M$, in gauge-mediated models, is of order
the messenger scale.  Assuming a value of this scale of order
the GUT scale, as in many models\cite{murayama,aharonyseiberg},
\beq
\delta m_{H_U}^2/M_Z^2 \approx 130;
\eeq
if $M= 10^2$ TeV, $130$ is reduced to $21$, still suggestive of a $5\%$ fine tuning.

A number of models have been proposed recently in which the predictions of MGM
do not hold\cite{dm2,othersquashed,shihetal}.   Meade, Shih and Seiberg\cite{mss}
have provided a general framework for considering gauge mediated models.
First, they give a definition:  a model is gauge mediated if the couplings between
the hidden and the visible sector vanish as the gauge couplings tend to zero.
As they stress, framing this definition raises interesting questions, especially
since, in a theory which is truly gauge-mediated in this sense,
the $\mu$ term cannot arise from supersymmetry-breaking dynamics.
In the course of this paper, we will return to this and other issues raised
by this definition.  Meade et al go on to characterize the most general
spectrum of gauginos and squarks and sleptons which can arise in this framework
(``General Gauge Mediation" or GGM).  Setting aside a possible Fayet-Iliopoulos
D term, there are, in general, six parameters which
characterize the low energy superparticle
spectrum.  We will discuss existing models,
some of their problems, and
their parameter counting.  We will then
extend these constructions, describing simple, weakly coupled theories
with up to the full set of six parameters.

Very often, in the metastable models with DSB, the low energy theory can be described by
a renormalizable theory without gauge interactions.  Motivated
by this, and by a desire for simplicity and explicitness, Shih and collaborators have
pursued a program of model building with weakly coupled fields, without gauge
interactions.  The discussion above suggests the following rules:
\begin{enumerate}
\item
The model should possess an R symmetry.
\item
There should be fields in the model with R charge not equal to zero or two.
\item
There should be a rich enough structure that the predictions of minimal gauge
mediation do not hold.
\end{enumerate}
Within this framework, the authors of \cite{shihetal} generate a class of models with
interesting features:  an MGM spectrum for gauginos, but not for squarks and
sleptons.  But the models suffer from certain difficulties.  Most important,
unless certain parameters are tuned, there are one loop corrections to squark
and slepton masses (squared) proportional to hypercharge.  These are problematic
parametrically, since gaugino {\it masses} are generated at one loop, and
because some masses will be tachyonic (in the absence of tuning).  One of the goals
of the present paper is to explore these issues.  We will consider ways to
``fix" the models of \cite{shihetal} so as to obtain, automatically, an
approximate symmetry (first noted in \cite{dinefischler}) which can
suppress these one loop contributions.  This symmetry has been dubbed
``messenger parity" in \cite{dimopoulos}.  We will also introduce simpler classes
of models which exhibit this feature automatically.

Another interesting feature of these constructions is the frequent appearance
of "runaway" directions, directions in field space where, classically,
the potential tends to zero for large fields.  In the models
of \cite{shihetal}, these directions are separated from the state of interest
by a barrier, but it is interesting to ask why these directions arise, and whether
they can be useful for model building.
We will see that the existence of runaway is common, and related to symmetries\cite{shihetal}.
The argument will immediately indicate how such behavior can be avoided.
In the models of \cite{shihetal}, classically, there is a (pseudo) moduli space,
separated, as we indicated, by a barrier from the runaway directions.  On the
moduli space, there is a point of unbroken R symmetry, and a Coleman-Weinberg
calculation is required to determine whether or not the symmetry
is broken.  In our
modified construction, there are branches of the moduli space on which the
R symmetry is everywhere broken.  A Coleman-Weinberg calculation is still required
to determine the decay constant of the R axion and the precise spectrum of the
model.

In the next section, we survey a number of existing approaches to model building.
We review the results of \cite{shih} concerning
spontaneous breaking of R symmetries in models with only gauge singlets,
and introduce the problem of messenger parity.
We explain, following \cite{dm2}, that a model with a single $5,\bar 5$ pair of messengers
and several singlets with
scalar and $F$ term vev's provides an example of GGM with two parameters describing both
the gaugino and sfermion spectrum, with
messenger parity automatic.  We show that a model with the structure $10
+ \overline{10}$ gives a richer parameter set, with the possibility
of significant compression of the sfermion spectrum.   We explain why, without
imposing extra symmetries, one can obtain at most five parameters
with this set of constructions, and describe simple
models (with symmetries) with the full complement of GGM parameters. All
of these models automatically possess
an approximate messenger parity.  In section \ref{squashing} we discuss the
spectra of these models from a phenomenological viewpoint.   In section \ref{rsymmetry},
we discuss issues connected with breaking the R symmetry.  We present simple models
with supersymmetry and R symmetry broken by multiple singlets, needed for the GGM models
described in section \ref{existingmodels}.
We explain why runaway directions are typical of models with fields with
R charge different than zero or two.  We provide a simple
example of a model where the R symmetry is
everywhere broken on a branch of the moduli space.

In section \ref{shihsmodels} we consider other approaches
to model building.  We discuss the model of ``Extraordinary Gauge
Mediation" of \cite{shihetal}  In this model, the presence
of (multiple) messengers plays a
crucial role in supersymmetry breaking.  We explain why, without additional
fields, one cannot obtain messenger parity as an accident in these theories.
We exhibit the minimal additional field content required, and discuss some
of the challenges to building a working model.  We then consider the class of
models
in which the R symmetry is broken classically, without
runaway behavior in the hidden sector.  Coupling these
to messengers allows realizations of GGM with CP conservation,
before coupling to the MSSM fields.  As a result, EdM's are
highly suppressed.  In the conclusions, we discuss
the possible phenomenological implications of these observations.
The difficulties of building models with the full parameter
set of GGM suggest that gauge mediation may make robust predictions
beyond those of \cite{mss}.   We also remark on some questions of definition:
we critique the definition of gauge mediation in \cite{mss}, as well as a
definition of {\it direct mediation} given in \cite{dm2}.

\section{Implementing GGM}
\label{existingmodels}

For our purposes, it will be convenient to parameterize the GGM spectrum
in terms of the gaugino masses, $m_\lambda$, $m_w$ and $m_b$ and
three parameters contributing to sfermion masses, $\Lambda_c^2$,
$\Lambda_w^2$ and $\Lambda_Y^2$.  In terms of these latter numbers, the
sfermion masses are given by:
\beq
\widetilde m^2 = 2
\left[
C_3\left({\alpha_3 \over 4 \pi}\right)^2 \Lambda_c^2
+C_2\left({\alpha_2\over 4 \pi}\right)^2 \Lambda_w^2
+{5 \over 3}{\left(Y\over2\right)^2}
\left({\alpha_1\over 4 \pi}\right)^2 \Lambda_Y^2\right]
\eeq
In the MGM, there is a simple relation between the various parameters; our goal
is to construct models in which these are independent.

\subsection{MGM and Messenger Parity}

The simple model of gauge mediation, eqn. [\ref{simplemgm}], has one particularly
attractive feature.  Without tuning of parameters, it automatically
has an approximate messenger parity symmetry, under which
\beq
q \leftrightarrow \tilde q ~~ \ell \leftrightarrow \tilde \ell~~~~V \rightarrow -V
\eeq
This symmetry is necessarily violated by couplings of the MSSM fields, but
the symmetry is good enough to
ensure that an expectation value for $\langle D_Y \rangle$ is only generated
at high loop order, so the usual two-loop contributions of eqn. [\ref{scalarsmgm}]
give the dominant contribution to scalar masses.  This cancellation is discussed in the
appendix.

This model is hardly complete.
One needs to add
some additional structure to account for the vev of the superfield $X$ (and,
of course, the $\mu$ term).
With our requirement that the model possess an R symmetry, in light of
Shih's theorem, one needs to add fields with unconventional R charges.  Before
doing this, we consider generalizations with more fields and messengers.

\subsection{Multiple Singlets:  Simple models of GGM}

A very simple generalization, mentioned in \cite{dm}, contains several singlets
and a single set of messengers:
\beq
W = X_i\left(\lambda^i_q q \tilde q + \lambda^i_\ell \ell\tilde \ell\right) + F_i X^i.
\label{multiplesinglets}
\eeq
This class of models, again, automatically exhibits a messenger-parity symmetry.
It yields a spectrum of gauginos, squarks and sleptons, however, which is
already different than that of minimal gauge mediation.  There are now
two parameters which describe the full spectrum (here $\langle
X_i\rangle = x_i + \theta^2 F_i$):
\beq
\Lambda_q = {\lambda_q^i F_i \over \lambda_q^j x_j}~~~
\Lambda_\ell = {\lambda_\ell^i F_i \over \lambda_\ell^j x_j}
\eeq
($i$ and $j$ summed).
The masses of the gluinos are given by
\beq
m_{\lambda} = {\alpha_3 \over 4 \pi} \Lambda_q~~~~
m_w = {\alpha_2 \over 4 \pi} \Lambda_{\ell}~~~~
m_b =  {\alpha_1 \over 4 \pi} \left [{2\over 3} \Lambda_q +
\Lambda_\ell \right ].
\label{gluinoformula}
\eeq
Similarly, for the squark and slepton masses we have:
\beq
\Lambda_c^2 = \Lambda_q^2;~~\Lambda_w^2 = \Lambda_{\ell}^2;~~\Lambda_Y^2
= \left ({2 \over 3} \Lambda_q^2 + \Lambda_{\ell}^2 \right ). 
\label{scalarsemgm}
\eeq

Note that in these models, the unified prediction for the gaugino mass is lost.
There is, in general, a relative phase between $\Lambda_q$ and
$\Lambda_\ell$, which leads, potentially, to edm's for quarks and leptons
at one loop.  The details depend on the origin of $\mu$.  This problem
will be common to many, but not all, of the models we discuss.

While more general than MGM, this model is described by two independent parameters, rather
than the six (not allowing for a Fayet-Iliopoulos term) permitted by the analysis
of \cite{mss}.
There are thus four mass relations, which hold at the messenger scale:
\beq
\left ({\alpha_3 \over 4 \pi} \right )^2\Lambda_c^2 = m_{\lambda}^2
\eeq
\beq
\left ({\alpha_2 \over 4 \pi} \right )^2\Lambda_{w}^2 = m_{w}^2
\eeq
\beq
{2 \over 3} \alpha_1 \alpha_2 m_\lambda + \alpha_1 \alpha_3 m_w
- \alpha_2 \alpha_3 m_b = 0
\eeq
and
\beq
{2 \over 3} \Lambda_c^2 + \Lambda_w^2 - \Lambda_Y^2=0.
\eeq
Note that the range of the low energy parameters (gaugino masses, $\Lambda_c^2$, etc.)
depend on the details of the microscopic model.  For example, it is easy to see that
if there are $N$ singlets, the ratio of $m_\lambda^2$ to $\Lambda_c^2$ is at most $N$.

Additional parameters arise if we slightly complicate the messenger sector.
If, for example, we replace the $5$ and $\bar 5$ by a $10$ and $\overline{10}$ ($Q,\bar Q,
\bar
U, U,\bar E, E$), then there are three independent parameters which describe the low
energy spectrum.  The resulting model:
\beq
W = X_i\left(y_i \bar Q Q + r_i \bar U U + s_i \bar E E\right)
\label{wthree}
\eeq
still, automatically, respects a messenger parity symmetry.
Here it is natural to define the three parameters:
\beq
\Lambda_Q = {y_i F_i \over y_j x_j}~~~
\Lambda_U = {r_i F_i \over r_j x_j}~~~
\Lambda_E = {s_i F_i \over s_j x_j}~~~.
\label{tenparameters}
\eeq
In terms of these, the low energy GGM parameters are:
\beq
m_{\lambda} = {\alpha_3 \over 4 \pi} \left ( 2 \Lambda_Q + \Lambda_U \right )
~~~
m_{w}={\alpha_2 \over 4 \pi} 3\Lambda_Q
~~~
m_{b}={\alpha_1 \over 4 \pi} \left(\frac{4}{3}\Lambda_Q+2\Lambda_E+\frac{8}{3}\Lambda_U\right)
\label{gauginosten}
\eeq
and
\beq
\Lambda_c^2 = 2\Lambda_Q^2+\Lambda_U^2
~~
\Lambda_w^2 =  3\Lambda_Q^2
~~
\Lambda_Y^2 =
\frac{4}{3}\Lambda_Q^2+2\Lambda_E^2+\frac{8}{3}\Lambda_U^2  
\label{sfermionsten}
\eeq
We will discuss
the low energy spectrum of the model in section \ref{squashing}, but
note that the presence of $\bar E E$ means that the masses
of the lightest sleptons are not correlated with the masses of doublets
or triplets.

Combining the models with $10$ and $\overline{10}$
and $5$ and $\bar 5$ (specifically adding
the superpotentials of \ref{multiplesinglets} and \ref{wthree}) yields a theory which
still possess a messenger parity and which now exhibits five of the GGM parameters.
There are now five parameters which characterize the full sparticle spectrum.
\beq
~~~~m_{\lambda} &=&{\alpha_3 \over 4 \pi} \left ( \Lambda_q + 2 \Lambda_Q + \Lambda_U \right ),
~m_{w}={\alpha_2 \over 4 \pi} \left( \Lambda_l + 3\Lambda_Q \right), \nonumber \\
~m_{b}&=&{\alpha_1 \over 4 \pi} \left( \frac{2}{3}\Lambda_q+\Lambda_l+\frac{4}{3}\Lambda_Q+2\Lambda_E+\frac{8}{3}\Lambda_U\right)
\label{gauginosfive}
\eeq
and
\beq
\Lambda_c^2 = \Lambda_q^2+2\Lambda_Q^2+\Lambda_U^2
~~
\Lambda_w^2 =  \Lambda_\ell^2+3\Lambda_Q^2
~~
\Lambda_Y^2 =
\Lambda_q^2+\Lambda_\ell^2+\frac{4}{3}\Lambda_Q^2+2\Lambda_E^2+\frac{8}{3}\Lambda_U^2  
\label{sfermionsfive}
\eeq
Again, in this case, since the six low energy parameters are described by five microscopic
ones, there is a sum rule.  In this case, the rule is more complicated.  It turns out
to be eighth order in the masses, with 294 terms.

Five is the largest number of parameters
one can obtain with the requirements:
\begin{enumerate}
\item  Automatic messenger parity (in all of the above models, none of the fields, $q,\bar q,
\ell, \bar \ell, Q, \bar Q, U,\bar U, E, \bar E$, can mix, due to the gauge quantum numbers).
\item  Complete $SU(5)$ multiplets.
\item  Perturbative unification:  multiplets such as the adjoint permit the full set of six
parameters, but unification is problematic due to the large, negative beta functions.
\item  No additional symmetries, beyond gauge symmetries, which distinguish the messengers.
\end{enumerate}

Relaxing the last requirement permits construction of models with the full set of six
parameters.  For example, consider a theory with two $5$ and $\bar 5$'s and a single
$10$ and $\bar 10$, coupled to at least three
singlets, where there are discrete symmetries which permit only the couplings:
\beq
\lambda_\ell^{a i} X_a \bar \ell_i\ell_i + \lambda_q^{a i} X_a \bar q_i q_i
+ \lambda_Q \bar Q Q + \lambda_U \bar U U + \lambda_E \bar E E
\eeq
Now there are four parameters associated with the two $5$'s,
\beq
\Lambda_q^i = {\lambda_q^{a i} F_a \over \lambda_q^{b i} x_b}
~~~
\Lambda_\ell^i = {\lambda_\ell^{a i} F_a \over \lambda_\ell^{b i} x_b}.
\eeq
while $\lambda_Q,\Lambda_U$ and $\Lambda_E$, are as in equation [\ref{tenparameters}].
The gaugino and sfermion masses are now as in eqns. [\ref{gauginosfive},\ref{sfermionsfive}], but with
$\Lambda_q \rightarrow \sum \Lambda_q^i$ in the gaugino formulas,
and $\Lambda_q^2 \rightarrow \sum (\Lambda_q^i)^2$ in the sfermion mass formulas.
The three gaugino masses and the three $\Lambda^2$ combinations
multiplying the different $\alpha_i$ in the sfermion masses are all independent for a total of six parameters.

Because of the large, non-asymptotically free beta functions for $SU(3)\times SU(2)
\times U(1)$, unification in these theories is a delicate matter.  If the messenger
scale is $10$'s of TeV, then the couplings become strong before the unification scale.
For higher messenger scale, the theory can remain perturbative.


The increasing complexity associated with theories with more parameters raises the possibility
that, if gauge mediation
is realized in nature, the underlying theory
generates only a subset of the full set of GGM parameters, leading to
additional predictions.  We will describe shortly how one can construct models
with multiple singlets with suitable vev's.  First, however, we consider some other
possible model building strategies.

Models with one singlet and multiple messengers do not
have difficulties with messenger parity, but they have the MGM spectrum.
Taking the messengers to be $5_i$ and $\bar 5_i$,
by separate unitary transformations of the fields one can write:
\beq
W = X \left(\lambda_a^q \bar q_a q_a  + \lambda_a^{\ell} \bar \ell_a \ell_a\right)
\eeq
This class of models still respects an approximate messenger parity symmetry.
However, the spectrum is that of MGM. This is also true if the messengers
are in the $\overline{10}$ and $10$ representations.

With multiple singlets and multiple messengers, with symmetries,
we saw that one can readily construct models with
automatic messenger parity.  But without symmetries,
the situation is more complicated.
We cannot, in general, bring the superpotential to a simple, diagonal form, but instead,
have:
\beq
W = X_i \left(\lambda_{ab}^i \tilde q_a q_b  +\dots\right).
\eeq
We can also allow mass terms (as in \cite{shihetal}), $m_{ab} \tilde q_a q_b + \dots$.
Here we encounter a serious difficulty:  there is no messenger parity symmetry,
unless the $\lambda^i$'s are each diagonal.  As a result,
there is a one loop contribution to $\langle D_Y \rangle$ (this is illustrated
in a simple case in the Appendix),
unless there are additional symmetries (we will see examples shortly).

\section{Squashing the Spectrum}
\label{squashing}

The five parameter model
has interesting phenomenological features.
For plausible values of the parameters, it exhibits a significant ``squashing"
or ``compression" of the spectrum.  In other words, the masses of the squarks,
and the $SU(2)$ singlet and doublet sleptons, can be quite close.  This can
appreciably ameliorate the fine tuning problems of gauge mediation, especially
if the scale of the messenger masses is low.
To illustrate in a simple limit,
take two of the five parameters to vanish:
\beq
\Lambda_Q =\Lambda_U = 0
\eeq
This leaves us with what we will call the ``three parameter model."($\Lambda_E$, $\Lambda_\ell$, and $\Lambda_q$).
The sparticle spectrum can be squashed in
two different ways; both exhibit interesting phenomenology.
First, one can take $\Lambda_\ell \sim \Lambda_E > \Lambda_q$.
In this case we can raise the slepton masses
so for example: $\tilde{m}_{\tilde{e}_R} \sim \tilde{m}_{sq}\sim 1\rm{TeV}$.
In \cite{shihetal} it was pointed out that this region of parameter space can allow a tuned cancelation between the soft Higgs mass generated by integrating out the messengers and the radiative corrections from the heavy stop mass. This allows a small $\mu$-term. The second way to squash the spectrum is by taking $\Lambda_\ell \sim \Lambda_E >\Lambda_q$ in such a way that $\tilde{m}_{sq} \sim 300 \rm{GeV}$ and $\tilde{m}_{\tilde{e}_R} \sim 100 \rm{GeV}$.
Such a limit
ameliorates the ``little hierarchy" problem mentioned in the introduction
which arises because heavy squarks typically renormalize the Higgs soft mass
to a large, negative value. However, even in this ``light" region of
parameter space the requirement that the lightest chargino mass be greater
than $100 \rm{GeV}$ implies that $m_{H_u}$ gets an $SU(2)$
charged messenger contribution: $m^{2}_{H_u(SU(2))} \geq (150 \rm{GeV})^2$.
The squarks then renormalize $m_{H_u}$ to give: $m^2_{H_u(sq)} \leq -(170\rm{GeV})^2$.
This still implies a modest amount of fine tuning.


\section{Breaking the R Symmetry}
\label{rsymmetry}

Shih\cite{shih} provided a simple model whose Coleman-Weinberg potential leads to breaking
of R symmetry.  The model has fields with R charges $1,-1,3$ and $2$, $\phi_1,\phi_{-1},
\phi_3, X$:
\beq
W = -FX + \lambda X \phi_1 \phi_{-1} +m_1 \phi_{1}^2+ m_2 \phi_{-1} \phi_{3}.
\label{shihsimple}
\eeq
The theory has a pseudomoduli space with $\phi_i = 0$ and
$X$ undetermined.  The Coleman-Weinberg analysis leads to
a non-zero value of $X$ at one loop.


We have seen that perhaps the simplest way to obtain a more general
gauge mediated spectrum is with multiple singlets with scalar and F term
vev's.  A simple example of such a model is provided by taking
several copies of the model of eqn. [\ref{shihsimple}]:
\beq
W = -FX^a + \lambda X^a \phi^a_1 \phi^a_{-1} + m_1^a \phi_1^a \phi_{1}^a
+ m_2^a \phi_{-1}^a \phi_3^a.
\label{multiplesinglet}
\eeq
Obviously, this is not the most general model consistent with symmetries;
one should allow couplings among the fields with different labels, $a$.
But given that this model has broken symmetries at a
(meta)stable local minimum, one sees that at least
for small values of these additional parameters,
one can obtain the desired structure:  multiple singlets,
broken supersymmetry, and broken $R$ symmetry.
The singlets can then be coupled to messengers, allowing the full
GGM spectrum.

While an existence proof, the existence of so many fields (a minimum of nine
in the above construction) is perhaps unappealing.  A model with fewer fields
exploits the couplings to messengers to fix the $R$-symmetry breaking vev's.
Consider
\beq
W = -F_a X_a + \lambda_a X_a \phi_1 \phi_{-1} +m_1 \phi_{1}^2+ m_2 \phi_{-1} \phi_{3}
+ y_a X_a\bar 5 5.
\label{lindasimpler}
\eeq
Here there are two $X$ type fields, but only a single set of $\phi$ fields.
For simplicity, we have indicated only a single set of $5$,$\bar 5$ messengers; the
generalization with $10$, $\overline{10}$ and/or multiple
$5$'s and $\bar 5$'s is immediate.  In this model, one linear combination
of $X$'s, call it $X_2$,  decouples from the $\phi$'s; the other, $X_1$,
obtains a vev from one loop diagrams
containing $\phi$ fields, as in the 
previous model.  $X_2$ then receives an (in general non-zero)
vev from loops of messengers.  For a suitable range of parameters, the messenger masses
are non-tachyonic.

We should note that a simpler version of eqn.[\ref{shihsimple}] omits the field $\phi_{-3}$
and the mass parameter $m_2$:
\beq
W = -FX + \lambda X \phi_1 \phi_{-1} + {m} \phi_1^2.
\eeq
This model has a runaway direction starting from the origin in field space.
Still, the Coleman-Weinberg analysis yields a local minimum of the potential
for $X$, with $\phi_1 = \phi_{-1} =0$.  For small $\lambda$, this minimum
is highly metastable.  While simpler, however, the range of parameters over
which the $R$ symmetry is broken is somewhat tuned. There is a metastable minimum for 
${\lambda F\over m^2}\lsim 3\cdot10^{-3}$. Approximating the potential with those considered in \cite{Duncan:1992ai} we find that this minimum is long lived in the limit $\lambda \left({F \over m^2}\right)^{1\over 2} \rightarrow 0$.

\subsection{Breaking R Symmetry at Tree Level:  The Problem of Runaway}

In the model of eqn.[\ref{shihsimple}], in addition to the pseudomoduli space,
one can obtain vanishing of the energy as a limiting process.
If
\beq
\lambda \phi_{-1} \phi_1 = F
\eeq
while, at the same time, $\phi_1 \rightarrow 0$, $\phi_X \rightarrow 0$,
then the potential tends to zero.  More precisely we can take:
\beq
\phi_1 = e^{\alpha} \sqrt{{F\over \lambda}};\; \phi_{-1} = e^{-\alpha} \sqrt{F\over \lambda};\;
 X = -{2 m_1 \over \lambda} e^{2 \alpha};\; \phi_3 = {2 m_1\over m_2}\sqrt{F\over \lambda} e^{3 \alpha}
\eeq
we see that
\beq
F_X=0;\; F_{\phi_1}=0;\;F_{\phi_{-1}}=0
\eeq
while
\beq
F_{\phi_{3}} = m_2 \sqrt{F \over \lambda} e^{-\alpha}
\eeq
so the potential tends to zero as $\alpha \rightarrow \infty$.
It is easy to understand why this happens.  In general, the manifold of classical
solutions of the supersymmetry equations is larger than implied by the symmetry group;
it is described by the {\it complexification} of the group.  Here we have solved
a subset of the equations, those for which the $F$ terms have charge $0$ or smaller.
The manifold of solutions of this subset of equations is still described by the complexified group.  Since all of the
non-vanishing $F_i$'s have R charge greater than zero, they vanish
as we take the parameter of the complexified group transformation to $\infty$.

Note that these field configurations, in which the R symmetry is everywhere broken,
are distinct from the pseudomoduli space.  As shown in \cite{shih}, the pseudomoduli
space is stabilized.  Decay to the runaway directions must proceed by tunneling.
Note that, because the runaway is classical, quantum effects, at weak coupling,
will not give rise to metastable minima in these directions.

This argument indicates that in theories with fields with non-standard R charges,
the appearance of runaway behavior is common\cite{shihetal}.  All that is required is that one be
able to solve the equations for the vanishing $F$ terms for all fields with R charge
greater than or equal to two, or less than or equal to two.  On the other hand, this
discussion also makes clear how one can avoid the runaway:  it must
not be possible to satisfy these
equations.  In this case, one has a branch
of the pseudomoduli space on which
the R symmetry is everywhere broken.  In the following subsection, we give an example
of such a model.

\subsection{Breaking R Symmetry at Tree Level:  No Runaway}

Based
on our discussion
above, the existence of runaway directions in models with fields with $R$
charge both greater than and less than two would seem to be a generic
feature.  Divide the fields of the model into a set with R charge greater than two,
$Y_i$, a set with $R$ charge less than two, $Z_i$, and a set with $R$ charge
equal to two, $X_i$.  (Note that with this division, the R charges of F components of
chiral superfields with $R < 2$ are negative.)  Suppose one can solve the equations
$\partial W /\partial \phi_i =0$ for $X_i$ and $Y_i$.  Then, even if we can't solve the
equations for $Z_i$, since all of the remaining $F$'s have
the $R$ charge of the same sign,
we can make the corresponding $F_i$'s arbitrarily small (while keeping the rest
zero) by a
complexified R transformation.  Clearly the same is true if we switch the role of $Y_i$ and
$Z_i$.  Roughly speaking, to determine if such solutions exist, we have to count
equations and unknowns.  Essentially we need, both for $R \geq2$ and $R\leq2$, an overdetermined
set of equations.

As an existence proof, consider a model with superpotential (the R charges
of the fields are indicated by the subscripts):
%
%
%
\begin{eqnarray}
W= X \left( \gamma \phi_{2/3} \phi_{-2/3} -\mu^2\right)+{\delta\over 3} \phi_{2/3}^3 +m_1
\phi_{2/3} Y_{4/3} +m_2 \phi_{-2/3} Y_{8/3}.
\end{eqnarray}
All of the parameters in the model can be taken to be real.

There is a branch of the (pseudo)moduli space with all fields $0$ except for $X$ where the potential is equal to $V=\mu^4$.
However, there is a second branch on which
\beq
\phi_{2\over 3}=m_2\sigma e^{i\theta};\; \phi_{-{2\over 3}} = {m_1 \sigma}e^{-i \theta}.
\eeq
where $\sigma$ is defined such that $\mu^2={m_1 m_2\over \gamma}(1+\sigma^2 \gamma^2)$.
This branch exists for $\sigma\in R$ and on it the value of the potential is
$$
V={m_1^2 m_2^2\over \gamma^2}\left(1+2\sigma^2 \gamma^2\right)\leq \mu^4.
$$
Therefore it is lower than the previous one and stable.

Unlike the runaways we have discussed above, one cannot now allow,
say, $\phi_{3/2}$ to be large
and $\phi_{-3/2}$ small at the same time.  The equations
for $\partial W \over \partial Y_i$
and $\partial W \over \partial X$ are incompatible. The (pseudo)moduli space is of
real dimension $3$.
Again, one should note that on this branch, there is no point at which the $R$ symmetry is restored.
A one loop Coleman-Weinberg calculation is required to determine the values of the $X$ and
$Y_i$ fields.

This model, however, has a serious deficiency.  The symmetries allow an additional coupling
\beq
\delta W = h~ Y_{4/3}^2 \phi_{-2/3}.
\eeq
Adding this coupling reintroduces the problem of runaway.  One can avoid this difficulty
by adding an additional field, $\chi_{2/3}$, and a $Z_2$ symmetry, under which all fields
but $X$ and $\chi_{2/3}$ are odd.  For the superpotential we take:
\begin{eqnarray}
W= X \left( \gamma \phi_{2/3} \phi_{-2/3} -\mu^2\right)+{\delta\over 3} \phi_{2/3}^2 \chi_{2/3} +m_1
\phi_{2/3} Y_{4/3} +m_2 \phi_{-2/3} Y_{8/3} + \lambda \chi_{2/3}^3.
\label{workingmodel}
\end{eqnarray}
These are all of the couplings permitted by the symmetries.  The features of the model
are similar to those we encountered above.  Supersymmetry is broken, and there
is a branch of the moduli space on which R symmetry is everywhere broken. Notice also that all the parameters appearing in [\ref{workingmodel}] can be made real by
field redefinitions.

\section{Other Approaches to Model Building}
\label{otherapproaches}

The models with multiple
singlets we have described in section \ref{existingmodels}, while providing a realization of GGM, hardly
exhaust the possibilities for model building.  In this section, we consider
some other approaches.

\subsection{Extraordinary Gauge Mediation}
\label{shihsmodels}

Ref. \cite{shihetal} presented models in which the messengers coupled
at tree level to the goldstino, and in which the dynamics of the
messenger fields figures crucially in determining the pattern of R symmetry
breaking.  We will take the messenger fields $\phi_i,\tilde \phi_i$, to fill
out a $5$ and $\bar 5$ representation of $SU(5)$.  Writing the lagrangian
in a schematic, $SU(5)$ invariant form:
\beq
W = \lambda_{ij} \phi_i \tilde \phi_j + m_{ij} \tilde \phi_i \phi_j + X F
\eeq
where $\lambda$ and $m$ are such that the theory has a continuous R symmetry.
For example, if $i=1,2$,
\beq
W = X \left(\lambda_1 \phi_1 \tilde \phi_1 + \lambda_2 \phi_2\tilde \phi_2\right)
+ m \phi_1 \tilde \phi_2.
\label{shihmessengers}
\eeq

To avoid one loop D terms, it is necessary that $\lambda_1 = \lambda_2$ to
a high degree of approximation.  If we make this assumption, then one
has a spectrum for scalars with two of the three parameters allowed by
GGM, but, perhaps surprisingly, with ``unification" relations for the
gaugino masses\cite{shihetal}.

Imposing that $\lambda_1=\lambda_2$, one has
an approximate messenger parity, insuring that
the $D$ term is generated only at high
orders.  We might try to understand
this equality of couplings as arising
from a symmetry which interchanges
the labels $1$ and $2$. The symmetry is violated
only by the mass term, and this violation is soft (e.g. corrections to the kinetic
terms for $\phi_1$ and $\phi_2$ are finite, and arise first at three loops).
This suggests that one can understand the model as arising from
an exact $1 \leftrightarrow 2$ symmetry which is spontaneously broken.

We can illustrate this possibility by introducing two fields, $\rho_+$ and $\rho_-$,
with couplings
\beq
X\left(\phi_1 \tilde \phi_1 + \phi_2 \tilde \phi_2\right) + X \mu^2
+\rho_+ \phi_1 \tilde \phi_2 + \rho_- \tilde \phi_1 \phi_2.
\eeq
If $\rho_+$ has an expectation value, and $\rho_- = 0$, then
the low energy theory has the structure of Shih's model.  The $Z_2$
symmetry interchanges $\rho_+$ and $\rho_-$, as well as interchanging
the labels ``$1$" and ``$2$".

In order that the vanishing of $\rho_-$ be natural, we would like the model
to possess a symmetry under which $ \rho_-$ transforms and
$\rho_+$ is neutral.
Because $\rho_+$ and $\rho_-$ carry different $R$ charges, this means that it is possible
to define two different $R$ symmetries, or alternatively, one $R$ symmetry and an ordinary,
global $U(1)$. Since these symmetries are spontaneously broken, there are additional
Goldstone bosons, and, unless $\phi_{\pm}$ couple to $X$ at tree level, additional
pseudomoduli.
To try and fix the values of $\rho_{\pm}$, one might try to write, for example, couplings:
\beq
Z\left(\rho_+ \chi_- + \rho_- \chi_+ -\mu^2\right) + Y \rho_+ \rho_-
\eeq
in the hopes of obtaining a pattern of vev's with the desired structure.
In this model, however, coupling of $Z$'s to $X$ cannot be forbidden, nor
certain additional dangerous couplings of $Y$.  Ignoring this, we encounter
the runaway issues which we have encountered before.

It is easy to achieve the desired structure in a theory with a classical moduli
space, if we are willing to introduce more fields and more symmetries.
E.g.
add
\beq
Y_1 (\rho_+ \chi_-) + Y_2 (\rho_- \chi_+)
\eeq
This model has an additional (ordinary) $U(1)$ symmetry under which the
fields $Y_i$ transform.  As a result, there are not additional dangerous couplings of $Y$
to $X$ (or $\chi$).  The model, as expected, has a moduli space with several branches.
There is a branch of the desired type, with $\rho_+ = a$, $\chi_+ = b$, and all other
fields vanishing.  On this branch, the only extra light fields are $\rho_+$ and
$\chi_+$.   Achieving a model where classically
one does not have additional moduli of this type
is difficult.
Whether in a complete model, the Coleman-Weinberg calculation can
yield a sensible metastable minimum along one of these branches is a question
which requires investigation.


\subsection{Broken R Symmetry at Tree Level, Multiple Messengers}

We have outlined above how to obtain a pseudomoduli space on which R symmetry
is broken everywhere.  We can easily write a model
of this type, coupled to messengers.
Start with the model of equation [\ref{workingmodel}].
As explained previously, this is a model where one cannot set all
of the $F$ terms with $R$ charge greater than or less than two to zero.
As a result, supersymmetry is broken, and there is a branch of the
classical moduli space in which the $R$ symmetry is everywhere broken.

Introduce couplings of the $Y$ fields to messengers:
\begin{eqnarray}
W_m= a Y_{4\over 3} (M_1 \overline {M}_1)+b Y_{8\over 3} (M_2 \overline{M}_2)
\end{eqnarray}
These couplings are schematic.  $M_i$'s can be $5$ or $10$'s of $SU(5)$, and
the Yukawa couplings need not be unified.  This is at most a two
parameter model, in which messenger parity holds automatically.  
This model has the virtue that CP is automatically conserved.  More complexity
is required, however, to obtain more than two parameters.

The messenger pair $M_1\overline{M_1}$ has R charge $2/3$ while the pair $M_2\overline{M_2}$
has R charge $-2/3$.  One needs to check, then, that the local minimum obtained from the
Coleman-Weinberg calculation without messengers does not yield tachyonic masses
for messengers.  This can be shown to hold for a range of parameters.

\section{Conclusions}
\label{conclusions}

We have presented entire classes of models with GGM spectra, and with
messenger parity as an automatic feature.  We have explored qualitatively different
possibilities for R symmetry breaking, in which there are branches of the pseudomoduli
space on which the R symmetry is everywhere broken.

The discussion of this paper shows that it is easy to construct
models with more parameters than that of the MGM, and that even
in weakly coupled theories, one can obtain the full
set of GGM parameters.  Models with more parameters are progressively
more complicated.  We have seen that one must be careful to insure the
existence of a messenger parity symmetry.  The fact that
obtaining the full set of GGM parameters
requires extra symmetry structure suggests that
models with few parameters might be more
likely.  We have indicated simple ways in which to compress the spectrum,
obtaining models without the severe fine tuning of MGM.  We have also
seen that problems with CP violation are typical of models in which
the MGM predictions are modified, but we have also seen exceptions.  The exceptions
arose in cases where there are several singlets, with distinct quantum
numbers under some symmetry.  An alternative solution to the problem
of edm's is that CP violation is spontaneous, as discussed in \cite{dm2,nir}.

One might wonder about our focus (and that of \cite{shihetal}) on renormalizable
theories with continuous R symmetries.  After all, we do not expect fundamental
theories to exhibit global, continuous R symmetries.  But the work of \cite{dm2}
on retrofitted model building suggests that it is challenging to build models
{\it without} an approximate R symmetry at the level of the low energy lagrangian.
This symmetry must then be spontaneously broken, and the criteria discussed by Shih
then must be satisfied.  In \cite{dm2}, low energy models without continuous symmetries
{\it were} constructed, but they required the existence of certain quite small
Yukawa couplings.  These, it was argued, could arise by a Frogatt-Nielsen mechanism.
One can debate whether these models are more or less complicated (or plausible) than
those presented in this paper.

These studies have lead us to rethink certain questions of definition.  We are not
entirely satisfied with the definition of gauge mediation presented in \cite{mss}.  In
particular, it can be applied to certain proposals, like anomaly mediation, in which
gravitational effects (or more generally, very high scale effects) are important.  In the
case of anomaly mediation, for example, if one tunes the Kahler potential to have
the sequestered form (or perhaps provides a higher dimensional, dynamical explanation), then
there are no contributions to scalar and gaugino masses in the limit that the
gauge couplings vanished.  Of course, the spectrum one obtains is problematic, and
the solutions to this problem might involve non-gauge interactions.  Still, it is not
clear whether one really wants to call even the unrealistic version ``gauge mediation".

Another question of definition has to do with the term ``direct mediation".  Loosely,
what is usually meant by this is that the messengers couple directly (as opposed to
through loop effects) to the fields responsible for the underlying breaking of
supersymmetry.  In \cite{dm2}, a definition was offered, that mediation is direct
if in the limit that the couplings to messengers are turned off, supersymmetry
is restored.  This definition is appealing.  Models which fit in
this framework include strongly coupled models with dynamical supersymmetry
breaking, in which some of the fields
of the strongly coupled sector
act as messengers.   But this definition is readily seen, after a moment's thought, to
be too restrictive.  Any O'Raifeartaigh model where the messengers couple to the
field $X$ with a non-vanishing F component, such
as that of eqn. [\ref{shihmessengers}] would not fit this definition.  A
more precise definition would be:  mediation is indirect if, decoupling
the messengers, the features of the hidden sector (spectra and patterns of R symmetry
breaking) are not appreciably affected.

While recent developments in dynamical supersymmetry breaking and gauge-mediated
model building may not make discovery of gauge mediation seem a certainty, they
do make the possibility much more plausible.  The phenomenology is likely
to be much richer than that of MGM, yet still restricted.  The detailed examination
of the possible parameter space is worthy of further exploration.

\noindent
{\bf Acknowledgements:}
We thank David Shih and Nathan
Seiberg for several valuable conversations and useful comments.  We thank
Nima Arkani-Hamed for discussions of general issues of tuning
in supersymmetry and gauge mediation, and for stressing the question
of CP violation in models of GGM.  This work
supported in part by the U.S. Department of Energy.

\noindent
{\bf Appendix}

\begin{figure} 
\includegraphics{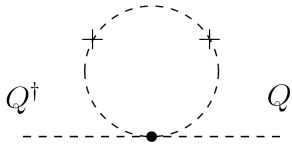} 
\caption{\label{diag1}1-loop Diagram for squark mass after integrating out the $U(1)_Y$ auxiliary field ``D".}
\end{figure}
\begin{figure}
\includegraphics{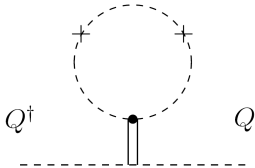} 
\caption{\label{diag2}1-loop Diagram for squark mass including $U(1)_Y$ auxiliary field ``D".}
\end{figure}

 To illustrate the problem of one loop D-terms, consider a model with a pair of messengers $\tilde{q}_i$ and $q_i$ ($i=1,2$). If the messengers both carry $U(1)_Y$ charge then the component lagrangian will contain quadratic interactions between the Messengers and the squarks and sleptons of the form
\beq \label{noD}
\mathcal{L}_{int} \sim g_1^2\left( \frac{Y_{q_i}}{2}\right) \left( \frac{Y_Q}{2}\right)q_i^{\dagger}q_iQ^{\dagger}Q.
\eeq
This can lead to the graph shown in \fig{diag1}.
In order for these contributions to be zero, it is sufficient that there exist a symmetry in the Messenger sector where  $\tilde{q}_i \leftrightarrow q_j$ and $V \rightarrow -V$. If such a symmetry exists, then it necessarily implies that no linear term in V (or any component of V) can be generated at one loop in the Lagrangian. The ''D-term" is an auxiliary component of V. A linear term in D is forbidden by this Messenger parity symmetry.  The absence of a linear term in D implies that the one loop graph for the squark mass must be zero. This is easily seen by not integrating out the D-term as is typically done. In this case, the one loop graph in \fig{diag1} is replaced by the graph in \fig{diag2}. So it is clear that if a linear term in D is forbidden then the one-loop squark mass is zero.


\begin{thebibliography}{9}


\bibitem{iss}
  K.~Intriligator, N.~Seiberg and D.~Shih,
  JHEP {\bf 0604}, 021 (2006)
  [arXiv:hep-th/0602239].

\bibitem{dfs}
  M.~Dine, J.~L.~Feng and E.~Silverstein,
  arXiv:hep-th/0608159.

\bibitem{dm}
  M.~Dine and J.~Mason,
  arXiv:hep-ph/0611312.


\bibitem{dinefischler}  
  M.~Dine and W.~Fischler,
  Phys.\ Lett.\  B {\bf 110}, 227 (1982).

\bibitem{earlymgm}
For a review of gauge mediation, with extensive references,
see  G.~F.~Giudice and R.~Rattazzi,
  Phys.\ Rept.\  {\bf 322}, 419 (1999)
  [arXiv:hep-ph/9801271].



\bibitem{dinenelson}
  M.~Dine, A.~E.~Nelson, Y.~Nir and Y.~Shirman,
  Phys.\ Rev.\  D {\bf 53}, 2658 (1996)
  [arXiv:hep-ph/9507378];
  M.~Dine, A.~E.~Nelson and Y.~Shirman,
  Phys.\ Rev.\  D {\bf 51}, 1362 (1995)
  [arXiv:hep-ph/9408384].


\bibitem{nelsonseiberg}
  A.~E.~Nelson and N.~Seiberg,
  Nucl.\ Phys.\  B {\bf 416}, 46 (1994)
  [arXiv:hep-ph/9309299].


\bibitem{shih}
  D.~Shih,
  arXiv:hep-th/0703196.


\bibitem{iss2}
  K.~Intriligator, N.~Seiberg and D.~Shih,
  JHEP {\bf 0707}, 017 (2007)
  [arXiv:hep-th/0703281].
  
\bibitem{katz}
  A.~Giveon, A.~Katz and Z.~Komargodski,
  arXiv:0804.1805[hep-th].
  



\bibitem{dm2}
  M.~Dine and J.~D.~Mason,
  arXiv:0712.1355 [hep-ph].


\bibitem{murayama}
  H.~Murayama and Y.~Nomura,
  Phys.\ Rev.\ Lett.\  {\bf 98}, 151803 (2007)
  [arXiv:hep-ph/0612186].

\bibitem{aharonyseiberg}
  O.~Aharony and N.~Seiberg,
  JHEP {\bf 0702}, 054 (2007)
  [arXiv:hep-ph/0612308].

\bibitem{othersquashed}
For early constructions with more general spectra, see, for example:
  S.~Dimopoulos, S.~D.~Thomas and J.~D.~Wells,
  Nucl.\ Phys.\  B {\bf 488}, 39 (1997)
  [arXiv:hep-ph/9609434].
  K.~I.~Izawa, Y.~Nomura, K.~Tobe and T.~Yanagida,
  Phys.\ Rev.\  D {\bf 56}, 2886 (1997)
  [arXiv:hep-ph/9705228].




\bibitem{shihetal}
  C.~Cheung, A.~L.~Fitzpatrick and D.~Shih,
  arXiv:0710.3585 [hep-ph].



\bibitem{mss}
  P.~Meade, N.~Seiberg and D.~Shih,
  arXiv:0801.3278 [hep-ph].





\bibitem{dimopoulos}
  S.~Dimopoulos and G.~F.~Giudice,
  Phys.\ Lett.\  B {\bf 393}, 72 (1997)
  [arXiv:hep-ph/9609344].









\bibitem{Duncan:1992ai}
  M.~J.~Duncan and L.~G.~Jensen,
  Phys.\ Lett.\  B {\bf 291}, 109 (1992).

\bibitem{nir}
  Y.~Nir and R.~Rattazzi,
  Phys.\ Lett.\  B {\bf 382}, 363 (1996)
  [arXiv:hep-ph/9603233].

\end{thebibliography}
\end{document}